\documentclass{article}
\usepackage{spconf,amsmath,graphicx}

\usepackage{booktabs} 

\title{A Study of Using Cepstrogram for Countermeasure Against Replay Attacks}
\name{Shih-Kuang Lee$^1$, Yu Tsao$^2$, Hsin-Min Wang$^1$}
\address{
$^1$Institute of Information Science, Academia Sinica, Taipei, Taiwan\\
$^2$Research Center for Information Technology Innovation, Academia Sinica, Taipei, Taiwan\\
\small\tt\{sklee, whm\}@iis.sinica.edu.tw, yu.tsao@citi.sinica.edu.tw}

\begin{document}

\maketitle

\begin{abstract}
This study investigated the cepstrogram properties and demonstrated their effectiveness as powerful countermeasures against replay attacks. A cepstrum analysis of replay attacks suggests that crucial information for anti-spoofing against replay attacks may be retained in the cepstrogram. When building countermeasures against replay attacks, experiments on the ASVspoof 2019 physical access database demonstrate that the cepstrogram is more effective than other features in both single and fusion systems. Our LCNN-based single and fusion systems with the cepstrogram feature outperformed the corresponding LCNN-based systems without the cepstrogram feature and several state-of-the-art single and fusion systems in the literature.
\end{abstract}


\section{Introduction}
\label{sec:intro}

Various speech-related applications use automatic speaker verification (ASV) to verify the identity of speakers. Despite the convenience it brings to people's lives, ASV systems are vulnerable to spoofing attacks \cite{Vulnerable1, Vulnerable2}, such as synthesized speech, converted speech, and replay attacks \cite{synthetic_speech, RawNet2, Res2Net}. Therefore, we require practical countermeasures against spoofing attacks.

Recently, a series of ASVspoof challenges have been conducted \cite{ASVspoof_2015, ASVspoof_2017, ASVspoof_2019, ASVspoof_2021}. The challenge encourages developing effective countermeasures to protect ASV systems from unforeseen spoofing attacks \cite{ASVspoof_2015}. It initially focused on synthesized speech \cite{ASVspoof_2015}, while trials of replay attacks were later added to the challenge of evaluating the performance of anti-spoofing \cite{ASVspoof_2017}. Because ASVspoof 2019 \cite{ASVspoof_2019}, the challenge is further divided into logical access (LA) and physical access (PA) scenarios.
The LA scenario includes synthesized and converted speech, and the PA scenario includes replay attacks. The anti-spoofing task has been the subject of growing research, and methods can be roughly divided into three types: front-end \cite{LFCC, LFCCN70, CQCC1, CQCC2, Zhang2021}, back-end \cite{synthetic_speech, RawNet2, Res2Net, STC2017, STC2019, wang2021}, and joint optimization of front-end and back-end \cite{tom18_interspeech, APSIPA19, Wu2020, Xie2021}. Front-end methods derive discriminative features, while back-end methods design effective models to distinguish between real and deceptive speech.

This study focused on the front-end processing for anti-spoofing purposes. We first analyzed the impact of replay attacks in the quefrency domain and investigated a common quefrency-domain representation of speech signals: the cepstrogram. The cepstrogram was obtained using the discrete cosine transform (DCT) to the spectrogram, which was obtained by applying the short-time Fourier transform (STFT) to the speech waveform.
Our investigations showed that the cepstrogram is more discriminative than the spectrogram when comparing real and deceptive speech data. We further examined the cepstrogram properties through experiments, which showed that the cepstrogram indeed carries key information for constructing countermeasures against replay attacks.

The remainder of this study is organized as follows. Section 2 reviews the related studies on anti-spoofing. Section 3 describes the proposed front-end method for countermeasures. Section 4 discusses the experimental setup and results. In section 5, we provide concluding remarks regarding this study.

\section{Related Work}
\label{sec:relate}
This section reviews related works, including common front-end and back-end methods for anti-spoofing tasks.

\subsection{Front-ends}

Front-end processing in anti-spoofing systems derives discriminative features for distinguishing between real and spoofed speech signals. This section reviews several speech features computed in the frequency and quefrency domains for anti-spoofing.

\subsubsection{Frequency-domain Features}

Spectrogram and CQTgram are frequency-domain features common in anti-spoofing systems \cite{Res2Net, STC2017, STC2019}. Spectrogram and CQTgram are obtained using STFT and constant Q transform (CQT), respectively, to speech signals \cite{CQCC1, CQCC2}. Phase information can be extracted from the spectrogram and CQTgram using several techniques such as the modified group delay function (MGD) and product spectrum (PS) to achieve high replay attack detection performance \cite{tom18_interspeech, APSIPA19, Groupdelay, prodspec2019}. However, state-of-the-art anti-spoofing results on recent datasets are obtained by directly using the magnitude representations of the frequency-domain features \cite{Res2Net}.

\subsubsection{Quefrency-domain Features}

Linear frequency cepstral coefficients (LFCC) \cite{LFCC, LFCCN70} and constant Q cepstral coefficients (CQCC) \cite{CQCC1, CQCC2} are quefrency-domain acoustic features commonly used in anti-spoofing systems \cite{Res2Net, ASVspoof_2017, ASVspoof_2019, STC2019}. The extraction of cepstral coefficients involves using cepstrum analysis of speech signals \cite{Oppenheim, DTSP}, where a set of linear filter banks is applied to the spectral features before transforming them into the quefrency domain. Although using linear filter banks may cause some loss of detail in the cepstrum, LFCC remains a popular quefrency-domain acoustic feature that achieves excellent performance in spoofed speech detection tasks \cite{Res2Net, STC2019}.

\begin{figure}[t]
  \centering
  \caption{The spectrogram (in log-scale) and cepstrogram (in log-scale) of the clean speech sample p262\underline{ }227 from VCTK \cite{VCTK}: (a) Spectrogram, (b) Cepstrogram, and (c) Cepstrum of the frame labeled by the yellow vertical line in (a) and (b).}
  \includegraphics[width=0.47\textwidth]{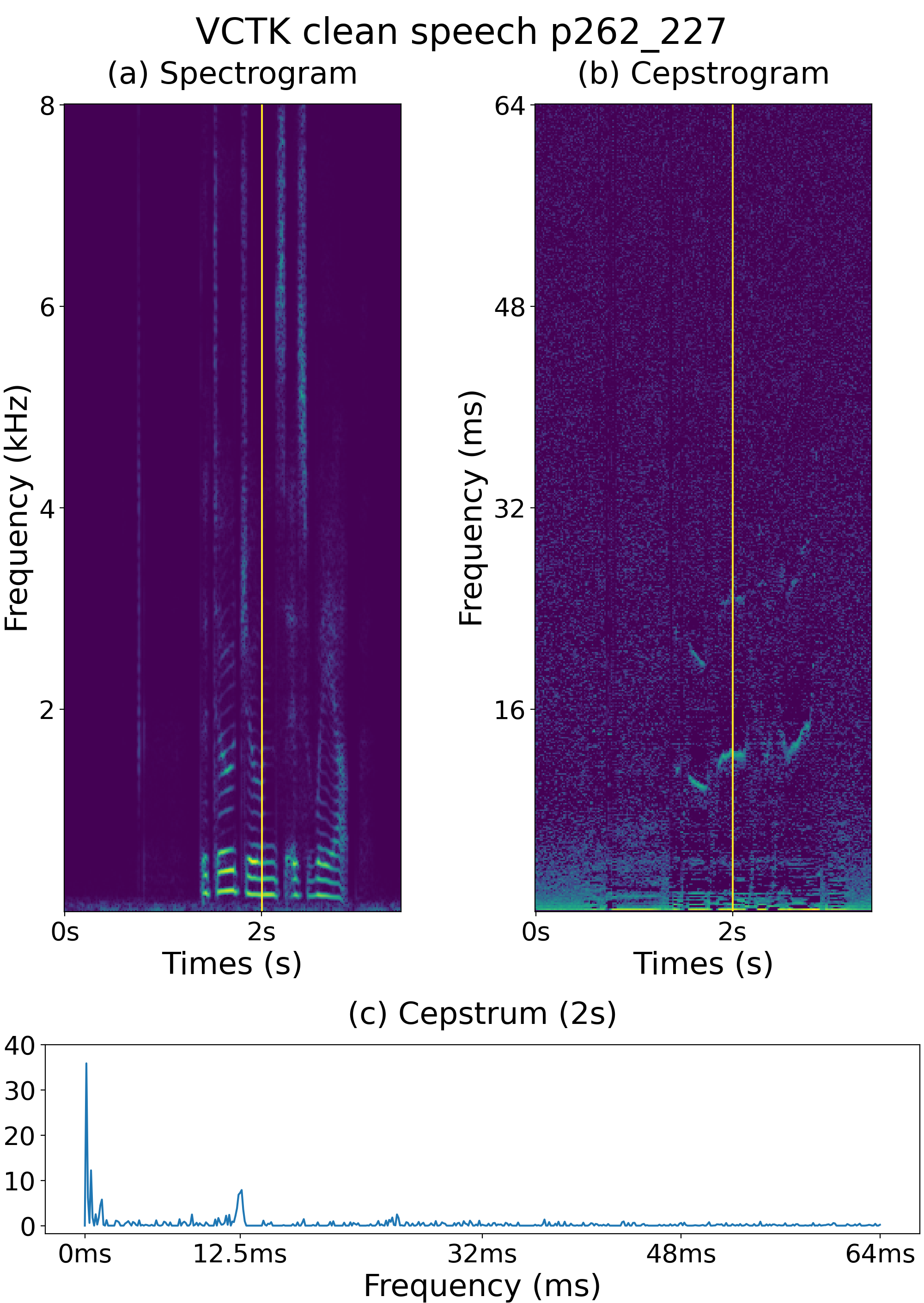}
  \label{fig:VCTK_clean}
\end{figure}

\begin{figure}[t]
  \centering
  \caption{The spectrogram (in log-scale) and cepstrogram (in log-scale) of the bona fide trial PA\underline{ }D\underline{ }0004063: (a) Spectrogram, (b) Cepstrogram, and (c) Cepstrum of the frame labeled by the yellow vertical line in (a) and (b).}
  \includegraphics[width=0.47\textwidth]{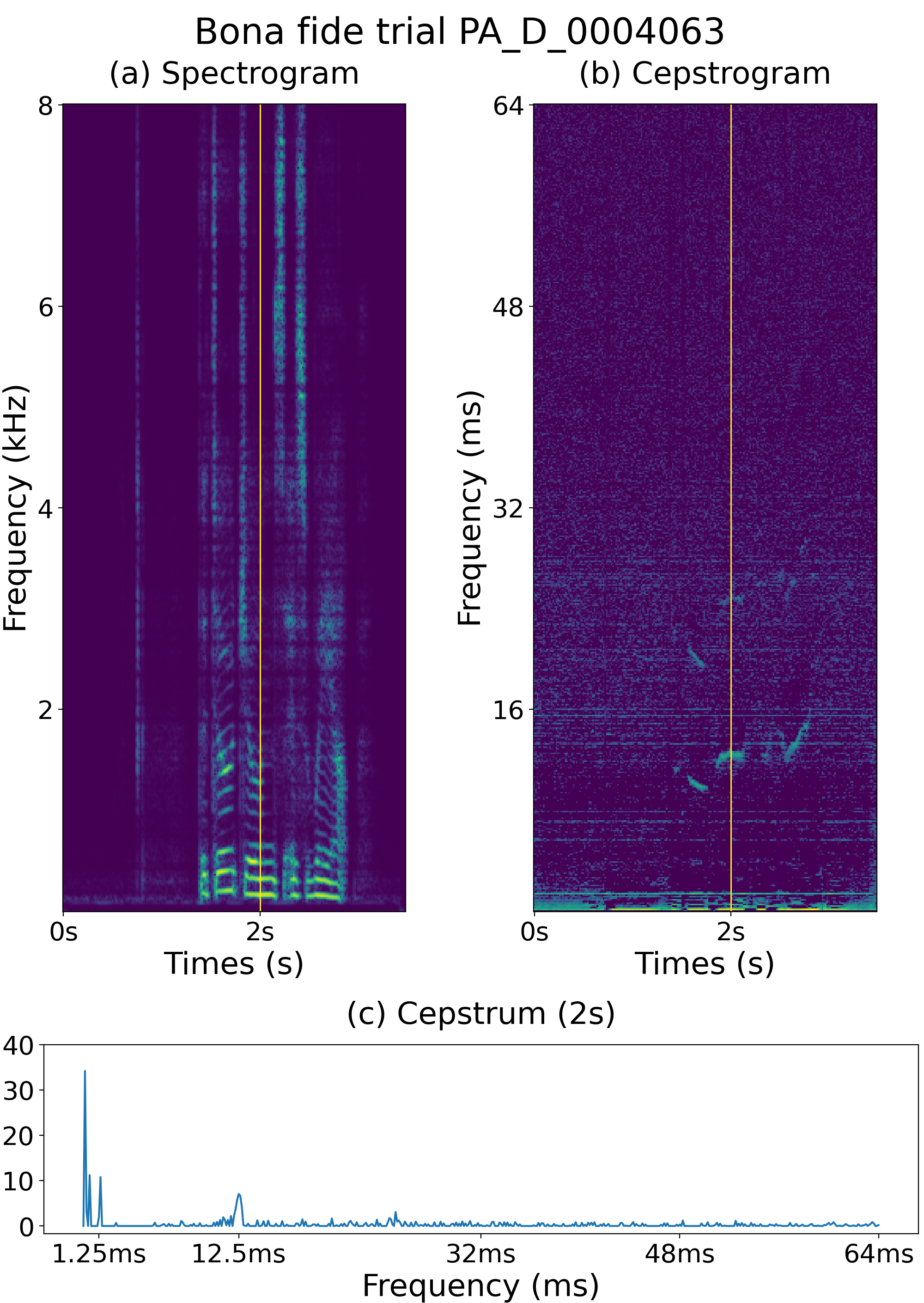}
  \label{fig:Bona_fide}
\end{figure}

\begin{figure}[t]
  \centering
  \caption{The spectrogram (in log-scale) and cepstrogram (in log-scale) of the spoofed trial PA\underline{ }D\underline{ }0024255: (a) Spectrogram, (b) Cepstrogram, and (c) Cepstrum of the frame labeled by the yellow vertical line in (a) and (b).}
  \includegraphics[width=0.47\textwidth]{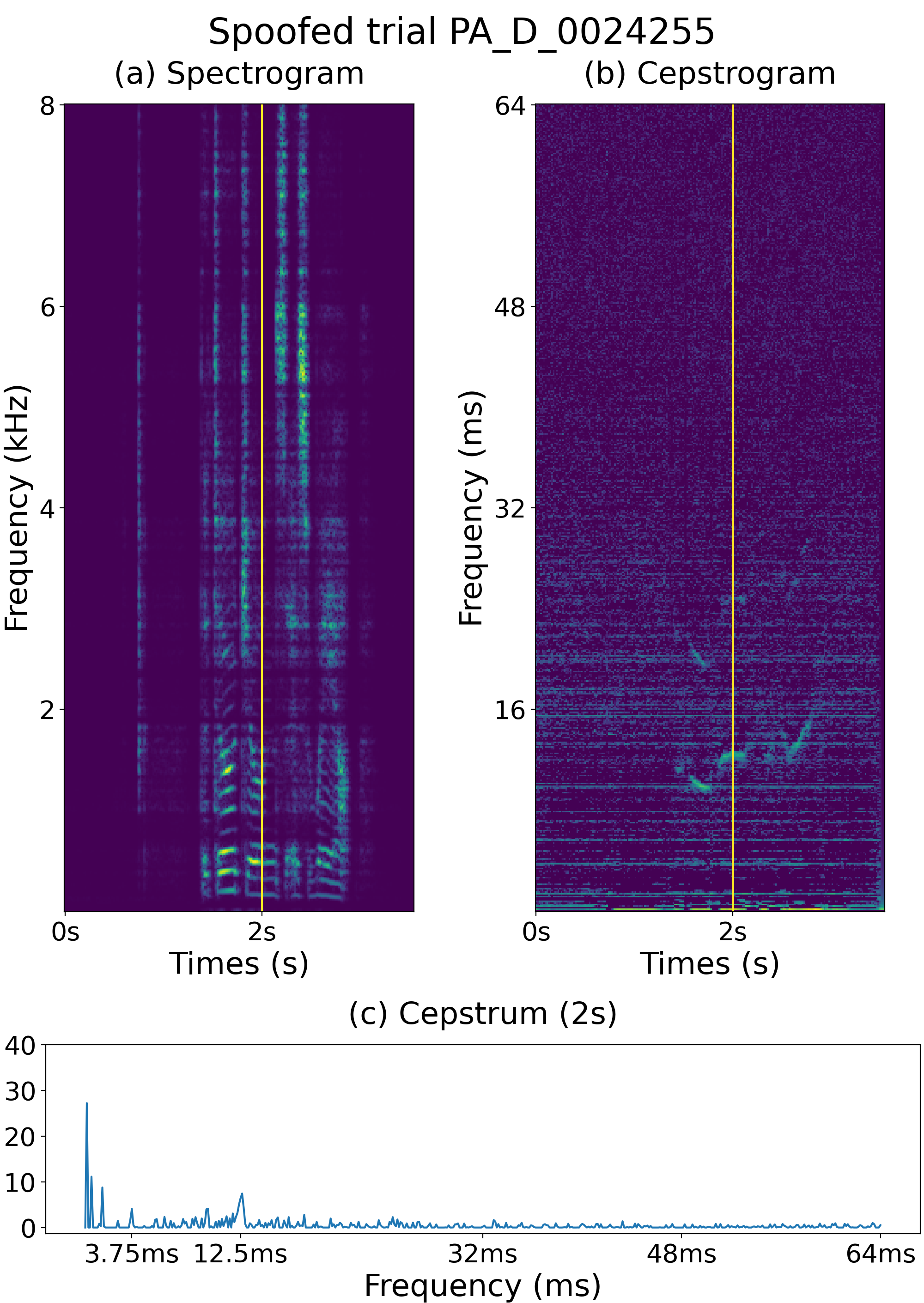}
  \label{fig:Spoof}
\end{figure}

\subsection{Back-ends}

The back end of the anti-spoofing system refers to different classification models. This section reviews two representative models: the Gaussian mixture model (GMM) and the deep neural network (DNN).

\subsubsection{Gaussian mixture model}

GMM-based classifiers are a classical method for countermeasures against replay attacks \cite{ASVspoof_2017, ASVspoof_2019, LFCC, LFCCN70} that determine whether a given speech signal is spoofed by calculating its likelihood. In such a system, one GMM calculates the likelihood that the speech is spoofed, and the other GMM calculates the likelihood that the speech is real. The two likelihoods are then compared to determine whether the speech is spoofed or real. Because the likelihood is calculated by averaging over all frames, the GMM does not exploit the temporal information of the speech features, resulting in limited detection performance.

\subsubsection{Deep neural network}

Compared with GMM-based anti-spoofing classifiers, DNN-based classifiers usually require only one model to detect spoofed speech \cite{synthetic_speech, Res2Net, STC2017, STC2019, tom18_interspeech, APSIPA19, ASVspoof_2019IEEE}. Most DNN-based classifiers use 2D convolutional layers in their model architecture to detect spoofing attacks with entire sequences of speech features, rather than single frames. In this manner, the spatiotemporal information in speech features can be more effectively characterized, thereby improving the countermeasure performance against replay attacks \cite{Res2Net, STC2019, tom18_interspeech, APSIPA19}.

\section{Cepstrogram}
\label{sec:ceps}
A cepstrogram is a time-quefrency representation of a speech signal, that is, its cepstrum changes gradually, while a spectrogram is a time-frequency representation of a speech signal, that is, its spectrum changes gradually. Fig.~\ref{fig:VCTK_clean} (a) and (b) present the spectrogram and cepstrogram, respectively, of a clean speech signal from the VCTK database \cite{VCTK}. For both plots, the vertical axis represents the time index, and the horizontal axis represents the frequency (for spectrogram) and quefrency (for cepstrogram). Colors represent signal amplitudes on the log1p (natural logarithm of (1 + input)) scale \cite{log1p, speechrecognition}.

Cepstrum is an effective representation that exploits homomorphic information of speech for various tasks, such as speech recognition \cite{speechrecognition}, speaker recognition \cite{speakerrecognition}, speech emotion recognition \cite{speechemotionrecognition}, pitch detection \cite{pitch_detection} and anti-spoofing \cite{CQCC1, CQCC2, STC2019, APSIPA19}. Further details on the properties of the cepstrum can be found in \cite{Oppenheim}.

To demonstrate the advantages of cepstrum analysis, we consider a scenario in which the signal involves an echo, in which the overall signal can be expressed as
\begin{equation}
x[n] = s[n] + {\alpha}s[n-{\tau}].
\label{eq1}
\end{equation}
The magnitude spectrum of the overall signal is
\begin{equation}
X(\omega) = S(\omega)(1 + {\alpha}^2 + 2{\alpha}cos({\omega\tau}))^{\frac{1}{2}}.
\label{eq2}
\end{equation}
After obtaining the logarithm of the magnitude spectrum, the overall spectrum can be further separated into two components
\begin{equation}
log(X(\omega)) = log(S(\omega)) + \frac{1}{2}log({1 + {\alpha}^2 + 2{\alpha}cos({\omega\tau})}).
\label{eq3}
\end{equation}

Based on Eq. (\ref{eq3}), we can view $log(X(\omega))$ as the original signal $log(S(\omega))$ with an additive periodic component ${\frac{1}{2}}log({1 + {\alpha}^2 + 2{\alpha}cos({\omega\tau})})$. By applying the DCT to $log(X(\omega))$, we obtain the cepstrogram of the signal $x[n]$. Periodic patterns can be easily observed in the cepstrogram, and therefore, it is argued that cepstrogram can be a powerful feature for replay attacks.

A signal with echoes exhibits rahmonic peaks in the quefrency domain, which reflects its pitch in the cepstrum analysis \cite{Oppenheim}. We expect speech with simulated reverberation or replay attacks to exhibit rahmonic peaks in the quefrency domain. Fig.~\ref{fig:Bona_fide} shows the spectrogram, cepstrogram, and cepstrum of a single frame of the bona fide trial generated from the speech sample shown in Fig.~\ref{fig:VCTK_clean} with the simulated reverberation, while Fig.~\ref{fig:Spoof} shows the spectrogram, cepstrogram, and cepstrum of a single frame of the spoofed trial generated from the bona fide trial speech in Fig.~\ref{fig:Bona_fide} for a replay attack. Comparing the cepstrogram of the bona fide trial shown in Fig.~\ref{fig:Bona_fide} (b) and that of the clean speech in Fig.~\ref{fig:VCTK_clean} (b), more rahmonic peaks can be observed in the former; for example, in addition to the largest peak at approximately 12.5 ms, a distinct peak at approximately 1.25 ms can also be observed, as shown in Fig.~\ref{fig:Bona_fide} (c). Comparing Fig.~\ref{fig:Spoof} (b) in Fig.~\ref{fig:Bona_fide} (b), the replay likely causes more peaks, for instance, there is an additional peak at approximately 3.75 ms in Fig.~\ref{fig:Spoof} (c). These figures confirm that whenever a speech signal is simulated with reverberation or replayed, additional rahmonic peaks appear in the quefrency domain. Accordingly,  a cepstrogram serves as a powerful feature against replay attacks, causing more harmonic peaks. The next section describes experiments conducted to confirm this conjecture.

\section{Experimental Setup and Results}
\label{sec:exps}
Experiments confirm that the cepstrogram is an effective feature for a replay attack spoofing task. Thus far, we experimented under the ASVspoof 2019 PA scenario and used the anti-spoofing system of the then-winning team T45 as the benchmark system to test the effectiveness of the cepstrogram feature. The ASVspoof 2019 PA task provides the largest database for replay attack detection \cite{ASVspoof_2019_database}. The T45 system performed best on the ASVspoof 2019 PA task with LFCC as one of the front-ends. Because we have front-ends other than LFCC that represent signals in different domains, our goal was to compare these front-ends with a fixed back-end. The results were evaluated in terms of the equal error rate (EER) and minimum tandem detection cost function (min-tDCF), which are standard metrics used in the ASVspoof 2019 challenge \cite{tDCF}.

\subsection{Experimental setup}

The T45 system uses a light convolutional neural network (LCNN) architecture as the back-end and the CQTgram, LFCC, and spectrogram via DCT (termed DCT) as the front-ends. Specifically, three single systems with different front-ends, namely LCNN-CQT, LCNN-LFCC, and LCNN-DCT, were first built on the training set. The development set determines the best models for testing the performance of the evaluation set. The scores from the three single systems are combined to generate the final score (for the fusion system). For a fair comparison, we followed the same setup as the T45 system to build the back end of our system while testing the new front end. In our experiments, CQTgram and LFCC were generated using the codes provided by the challenge organizer, and the other features were extracted using our codes.
Table~\ref{tab:team_T45} presents the performance of the T45 system reported in \cite{ASVspoof_2019, STC2019} (upper four rows) and the proposed implementation (lower four rows). The table shows that our implementations yield slightly better performance than those reported in \cite{ASVspoof_2019, STC2019}. This might be due to the dropout layer setting. The setting of the dropout layers of the T45 system was not specified in the corresponding study \cite{STC2019}, and we only performed one dropout on the flatten layer to prevent overfitting. For more details on our implementation, please visit our repository \footnote{https://github.com/shihkuanglee/RD-LCNN}.
In the following discussion, we use our implementation for comparison.

\subsection{Results}

Table~\ref{tab:LCNN_Spec_Ceps} lists the results of the systems using the LCNN as the back-end with five different front-ends: CQT, LFCC, DCT, Spec (spectrogram), and Ceps (cepstrogram). It is evident from the table that Ceps performs the best, confirming its effectiveness. Next, we incorporate the single system LCNN-Ceps into the fusion systems. As shown in Table~\ref{tab:Ceps_performance}, all fusion systems with the cepstrogram front-end performed better than their counterparts without the cepstrogram front-end. For example, the system using Ceps+CQT reduces the EER from 0.514 to 0.149 compared to the system using CQT, while the system using Ceps+LFCC+CQT+Spec reduces the EER from 0.177 to 0.094 compared to the system using LFCC+CQT+Spec. 

Finally, we compared our systems with the proposed Ceps feature with the state-of-the-art systems. The top and bottom panels in Table~\ref{tab:SOTA_19} show the results for the single and fusion systems, respectively. It is evident from the table that our system performs best in both single-system and fusion-system categories with the cepstrogram as the front-end. Our fusion system using CQT+LFCC+Spec (system LCNN-CQT+LFCC+Spec) achieves a 38\%\ (0.287 to 0.177) EER reduction on the evaluation set compared to the best fusion system that uses the same features (Res2Net-CQT+LFCC+Spec \cite{Res2Net}); the fusion system using CQT+LFCC+Spec+Ceps (system LCNN-CQT+LFCC+Spec+Ceps) can further reduce the EER to 0.094, which is the new state-of-the-art performance on the evaluation set of the ASVspoof 2019 PA task.

\begin{table}[ht]
\scriptsize
\centering
\caption{Performance comparison of the T45 system (reported in \cite{STC2019}) and our implementation. All systems used the LCNN architecture as the back-end.}
\label{tab:team_T45}
\begin{tabular}{rllll}
\\
\toprule
\multicolumn{1}{c}{} & \multicolumn{2}{c}{\textbf{Dev}} & \multicolumn{2}{c}{\textbf{Eval}} \\
\cmidrule(l){2-5}
\multicolumn{1}{c}{\textbf{System}} & \multicolumn{1}{c}{\textbf{tDCF}} & \multicolumn{1}{c}{\textbf{EER}}    & \multicolumn{1}{c}{\textbf{tDCF}} & \multicolumn{1}{c}{\textbf{EER}} \\
\midrule
\multicolumn{1}{l}{\textbf{CQT}\cite{STC2019}}  & 0.0197 & 0.800  & 0.0295  & 1.23 \\
\multicolumn{1}{l}{\textbf{LFCC}\cite{STC2019}} & 0.0320 & 1.311  & 0.1053  & 4.60 \\
\multicolumn{1}{l}{\textbf{DCT}\cite{STC2019}}  & 0.0732 & 3.850  & 0.560   & 2.06 \\
\multicolumn{1}{c}{\textbf{Fusion}\cite{STC2019}} & \textbf{0.0001} & \textbf{0.0154} & \textbf{0.0122} & \textbf{0.54}\\
\midrule
\multicolumn{1}{l}{\textbf{CQT}}  & 0.0096 & 0.374 & 0.0130 & 0.514 \\
\multicolumn{1}{l}{\textbf{LFCC}} & 0.0145 & 0.519 & 0.0299 & 1.061 \\
\multicolumn{1}{l}{\textbf{DCT}}  & 0.0385 & 1.444 & 0.0774 & 2.897 \\
\multicolumn{1}{l}{\textbf{Fusion}} & \textbf{0.0014} & \textbf{0.057} & \textbf{0.0048} & \textbf{0.165} \\
\bottomrule
\end{tabular}
\end{table}

\begin{table}[ht]
\scriptsize
\centering
\caption{Results of the single systems. All systems used the LCNN architecture as the back-end.}
\label{tab:LCNN_Spec_Ceps}
\begin{tabular}{rllll}
\\
\toprule
\multicolumn{1}{c}{} & \multicolumn{2}{c}{\textbf{Dev}} & \multicolumn{2}{c}{\textbf{Eval}} \\
\cmidrule(l){2-5}
\multicolumn{1}{c}{\textbf{System}} & \multicolumn{1}{c}{\textbf{tDCF}} & \multicolumn{1}{c}{\textbf{EER}}    & \multicolumn{1}{c}{\textbf{tDCF}} & \multicolumn{1}{c}{\textbf{EER}} \\
\midrule
\multicolumn{1}{l}{\textbf{CQT}}  & 0.0096 & 0.374 & 0.0130 & 0.514 \\
\multicolumn{1}{l}{\textbf{LFCC}} & 0.0145 & 0.519 & 0.0299 & 1.061 \\
\multicolumn{1}{l}{\textbf{DCT}}  & 0.0385 & 1.444 & 0.0774 & 2.897 \\
\multicolumn{1}{l}{\textbf{Spec}} & 0.0148 & 0.556 & 0.0522 & 1.719 \\
\multicolumn{1}{l}{\textbf{Ceps}} & \textbf{0.0039} & \textbf{0.129} & \textbf{0.0105} & \textbf{0.370} \\
\bottomrule
\end{tabular}
\end{table}

\begin{table}[ht]
\scriptsize
\centering
\caption{Results of the fusion systems.}
\label{tab:Ceps_performance}
\begin{tabular}{rllll}
\\
\toprule
\multicolumn{1}{c}{} & \multicolumn{2}{c}{\textbf{Dev}} & \multicolumn{2}{c}{\textbf{Eval}} \\
\cmidrule(l){2-5}
\multicolumn{1}{c}{\textbf{System}} & \multicolumn{1}{c}{\textbf{tDCF}} & \multicolumn{1}{c}{\textbf{EER}}    & \multicolumn{1}{c}{\textbf{tDCF}} & \multicolumn{1}{c}{\textbf{EER}} \\
\midrule
\textbf{     CQT} & 0.0096 & 0.374 & 0.0130 & 0.514 \\
\textbf{Ceps+CQT}  & \textbf{0.0024} & \textbf{0.094} & \textbf{0.0043} & \textbf{0.149} \\
\cmidrule(l){2-5}
\textbf{     LFCC} & 0.0145 & 0.519 & 0.0299 & 1.061 \\
\textbf{Ceps+LFCC} & \textbf{0.0030} & \textbf{0.109} & \textbf{0.0074} & \textbf{0.254} \\
\cmidrule(l){2-5}
\textbf{     DCT}  & 0.0385 & 1.444 & 0.0774 & 2.897 \\
\textbf{Ceps+DCT}  & \textbf{0.0013} & \textbf{0.074} & \textbf{0.0066} & \textbf{0.242} \\
\midrule
\textbf{     CQT+LFCC} & 0.0037 & 0.166 & 0.0079 & 0.283 \\
\textbf{Ceps+CQT+LFCC} & \textbf{0.0022} & \textbf{0.074} & \textbf{0.0042} & \textbf{0.150} \\
\cmidrule(l){2-5}
\textbf{     CQT+DCT} & 0.0048 & 0.205 & 0.0111 & 0.475 \\
\textbf{Ceps+CQT+DCT} & \textbf{0.0015} & \textbf{0.059} & \textbf{0.0034} & \textbf{0.128} \\
\cmidrule(l){2-5}
\textbf{     LFCC+DCT} & 0.0042 & 0.183 & 0.0089 & 0.321 \\
\textbf{Ceps+LFCC+DCT} & \textbf{0.0012} & \textbf{0.039} & \textbf{0.0040} & \textbf{0.143} \\
\midrule
\textbf{     LFCC+CQT+DCT} & 0.0014 & 0.057 & 0.0048 & 0.165 \\
\textbf{Ceps+LFCC+CQT+DCT}  & \textbf{0.0008} & \textbf{0.052} & \textbf{0.0029} & \textbf{0.104} \\
\cmidrule(l){2-5}
\textbf{     LFCC+CQT+Spec} & 0.0009 & 0.039 & 0.0045 & 0.177 \\
\textbf{Ceps+LFCC+CQT+Spec} & \textbf{0.0004} & \textbf{0.022} & \textbf{0.0027} & \textbf{0.094} \\
\bottomrule
\end{tabular}
\end{table}

\begin{table}[ht]
\scriptsize
\centering
\caption{Performance comparison of the Ceps-based system and the SOTA systems.}
\label{tab:SOTA_19}
\begin{tabular}{@{}lllll@{}}
\\
\toprule
\multicolumn{1}{c}{} & \multicolumn{2}{c}{\textbf{Dev}} & \multicolumn{2}{c}{\textbf{Eval}} \\
\cmidrule(l){2-5}
\multicolumn{1}{c}{\textbf{Single System}} & \multicolumn{1}{c}{\textbf{tDCF}} & \multicolumn{1}{c}{\textbf{EER}}           & \multicolumn{1}{c}{\textbf{tDCF}} & \multicolumn{1}{c}{\textbf{EER}} \\
\midrule
\textbf{T45-LCNN-CQT}\cite{STC2019} & 0.0197 & 0.800  & 0.0295  & 1.23 \\
\textbf{T28-ResNeWt-MFbank+CQT}\cite{APSIPA19} & 0.0093 & 0.41 & 0.0134 & 0.52 \\
\textbf{LCNN-CQT} & 0.0096 & 0.375 & 0.0130 & 0.514 \\
\textbf{Res2Net-CQT}\cite{Res2Net} & 0.0086 & 0.329 & 0.0116 & 0.459 \\
\textbf{LCNN-Ceps} & \textbf{0.0039} & \textbf{0.129} & \textbf{0.0105} & \textbf{0.370} \\
\midrule
\midrule
\multicolumn{1}{c}{} & \multicolumn{2}{c}{\textbf{Dev}} & \multicolumn{2}{c}{\textbf{Eval}} \\
\cmidrule(l){2-5}
\multicolumn{1}{c}{\textbf{Fusion System}} & \multicolumn{1}{c}{\textbf{tDCF}} & \multicolumn{1}{c}{\textbf{EER}}           & \multicolumn{1}{c}{\textbf{tDCF}} & \multicolumn{1}{c}{\textbf{EER}} \\
\midrule
\textbf{T45-Fusion}\cite{STC2019} & \textbf{0.0001} & \textbf{0.0154} & 0.0122 & 0.54 \\
\textbf{T28-Fusion}\cite{APSIPA19} & 0.0049 & 0.20 & 0.0096 & 0.39 \\
\textbf{Res2Net-CQT+LFCC+Spec}\cite{Res2Net} & 0.0028 & 0.096 & 0.0075 & 0.287 \\
\textbf{LCNN-CQT+LFCC+Spec} & 0.0009 & 0.039 & 0.0045 & 0.177 \\
\textbf{LCNN-CQT+LFCC+DCT} & 0.0014 & 0.057 & 0.0048 & 0.165 \\
\textbf{LCNN-CQT+LFCC+DCT+Ceps} & 0.0008 & 0.052 & 0.0029 & 0.104 \\
\textbf{LCNN-CQT+LFCC+Spec+Ceps} & 0.0004 & 0.022 & \textbf{0.0027} & \textbf{0.094} \\
\bottomrule
\end{tabular}
\end{table}

\section{Discussion}
\label{sec:disc}
The experiments confirm that the cepstrogram provides key information for anti-spoofing against replay attacks. As described in \cite{Oppenheim}, a signal with echoes exhibits rahmonic peaks in the quefrency domain. This inspired us to investigate whether the replayed speech also had rahmonic peaks in the cepstrum. From Fig.~\ref{fig:Spoof} (b), additional rahmonic peaks can be observed in the replay attack trial. For the LFCC feature, the spectral envelope became blurred owing to filter banks, causing such peaks to disappear. As shown in Table~\ref{tab:LCNN_Spec_Ceps}, the Ceps-based single system outperforms single systems using other features, including LFCC. As shown in Table~\ref{tab:Ceps_performance}, a fusion system using the cepstrogram feature constantly outperforms its counterpart without using the cepstrogram feature. The fusion system of LFCC-LCNN and LCNN-Ceps also results in performance improvement in contrast to single systems, although both are quefrency-domain features, and we leave it as future work and focus on the properties of the cepstrogram in this study.

\section{Conclusions}
\label{sec:conc}
This qualitatively analyzed the properties of the cepstrogram, a time-quefrency representation of a speech signal, and then applied it to anti-spoofing against replay attacks. The analysis shows that the cepstrogram shows additional rahmonic peaks associated with the specific echo patterns and thus serves as a powerful feature for replay attack detection. Our extensive experiments on the ASVspoof 2019 PA task also quantitatively confirmed the effectiveness of the cepstrogram against replay attacks. Using the cepstrogram as a front-end, our systems achieved state-of-the-art performance in both the single-system and fusion-system categories.






\bibliographystyle{IEEEbib}
\bibliography{mybib}

\end{document}